\documentclass{article}

\usepackage{arxiv}

\usepackage[utf8]{inputenc} 
\usepackage[T1]{fontenc}    
\usepackage{hyperref}       
\usepackage{url}            
\usepackage{booktabs}       
\usepackage{amsfonts}       
\usepackage{nicefrac}       
\usepackage{microtype}      
\usepackage{natbib}
\usepackage{lipsum}
\usepackage{graphicx}
\graphicspath{ {./images/} }

\title{Forecasting megaelectron-volt electron flux in the Earth’s outer radiation belt using supervised machine learning algorithms and time series foundation model}

\author{
 Rungployphan Kieokaew \\
  Univ Toulouse, CNES, CNRS, \\
  IRAP, Toulouse;\\
  INRIA, Centre Inria Paris;\\
  Augura Space, Paris, France\\
  \texttt{rungployphan.kieokaew@utoulouse.fr} \\
   \And
 Ryad Guezzi \\
  CentraleSupélec; \\
  INRIA, Centre Inria Paris;\\
  Augura Space, Paris, France\\
  \texttt{guezzi.ryad@gmail.com} \\
  \And
 François Ginisty \\
  INRIA, Centre Inria Paris;\\
  Augura Space, Paris, France\\
  \texttt{francois.ginisty@auguraspace.com} \\
   \And
 Hadrien Mariaccia \\
  Augura Space, Paris, France\\
  \texttt{hadrien.mariaccia@auguraspace.com} \\
}

\begin{document}
\maketitle
\begin{abstract}
Accurate forecasting of megaelectron volt (MeV) electrons in the outer Earth’s radiation belt, which can pose significant risks to satellites, is essential for risk mitigation and spacecraft operations. We develop a machine-learning-based pipeline for forecasting 1-MeV electron flux variations, focusing first on a 6-hour forecast horizon. Using precipitating electrons measured by POES NOAA-15, near 1-MeV electron flux measured by GOES, solar wind measurements near L1, and geomagnetic activity indices as inputs in 2013 – 2023, we train algorithms including linear regression, 1-D convolutional and long-short term memory neural networks, and Transformer-Encoder to forecast 1-MeV electron flux in McIlwain's L-shells between 2.8 and 6.0 with 0.1 bin resolution. Particularly, we exploit the time series foundation model “TimesFM” for (1) a zero-shot prediction and (2) a hybrid application involving the ridge regression on the past dynamic covariates combined with the TimesFM inference on the residuals. Using data in January-June 2024 as an out-of-sample test, we find that the hybrid application of TimesFM named “TimesFM+Cov” yields the best results with an average $R^2$ score of 0.9 over all L-shells, compared to the average $R^2$ under 0.78 of all other models. The $R^2$ of TimesFM+Cov remains above 0.9 for the L-shells between 2.8 and 4.7, and it drops to 0.77 at L=6.0, showing the improvement of 12\% at the lowest L-shell and of 48\% at the highest L-shell compared to our second-best models. Our work offers an alternative perspective on how a pre-trained foundation model could be adapted for space weather forecasting.
\end{abstract}


\section{Introduction}
Man-made satellites in medium- and high-altitudes are continuously exposed to hazardous space radiation in the Earth's outer Van Allen belt ($\sim$3-8 Earth radii). A primary threat is the relativistic electron population --- with energies reaching the megaelectron-volt (MeV) --- trapped within the outer Van Allen radiation belt \cite{https://doi.org/10.1029/JZ064i011p01683}. Due to their high penetration capability, these electrons easily bypass standard shielding \cite{2025SpWea..2304226H}. Depending on the solar activity and consequent electron injections, the electron intensities across the outer belt are enhanced \cite{https://doi.org/10.1029/2025SW004448}, sometimes increased by a factor of 10$^4$, e.g., the 2003 Halloween Storm \cite{2004Natur.432..878B}. During such peak intensity events, the MeV electron population can induce deep-dielectric charging and discharging in electronic systems, potentially leading to spacecraft anomalies and failures \cite{LAI2018419}. Mitigating the impact of these MeV electron events is thus a critical priority for the space industry, service providers, and government agencies.

Developing operational, real-time prediction of MeV electrons in the outer belt is critical for anticipating radiation effects and minimize radiation hazard risks. Using the Relativistic Electron Forecast Model \cite{1990JGR....9515133B}, the National Oceanic and Atmospheric Administration (NOAA)’s Space Weather Prediction Center (SWPC) provides 1, 2, and 3-day forecasts of $>$2 MeV electron fluxes at Geosynchronous orbit (GEO), utilizing a linear prediction filter driven by real-time solar wind data from the Advanced Composition Explorer (ACE) spacecraft at the Lagrange L1 point \cite{1998SSRv...86..257C}. For orbits below GEO, physics-based models have been adapted for real-time forecasting. The Satellite Risk Prediction and Radiation Forecast System (SaRIF) \cite{https://doi.org/10.1029/2021SW002822} was funded by the European Space Agency (ESA) Space Safety Programme. It provides 1-day forecasts of $>$800 keV and $>$2 MeV electron fluxes using real-time Geostationary Operational Environmental Satellite (GOES) data; however, these forecasts have been unavailable at the time of this writing since the decommissioning of GOES-16 on 7 April 2025. Furthermore, the ESA-funded Radiation Belt Forecast And Nowcast activity (RB-FAN) framework \cite{2024cosp...45.2851M} predicts various energies of electrons including at $>$1 MeV and $>$2 MeV up to 3 days in advance using the advanced solar wind prediction from the EUropean Heliospheric FORecasting Information Asset (EUHFORIA) \cite{refId0}. These modeling frameworks represent the international efforts for operational forecasting of high-energy radiation belt environments.

More recently, there have been developments of the MeV-electron forecasting in the outer radiation belt using linear regression and machine learning techniques. Using multivariate autoregression, \cite{2015SpWea..13..853S} proposed prediction models of the MeV electron flux variation up to 2 days in advance for the McIlwain L values between 3 and 6, and at GEO. To develop their models, the best set of explanatory parameters consisting of solar wind parameters, geomagnetic indices, and the electron log-flux at GEO were determined using the Akaike Information Criterion \cite{1974ITAC...19..716A} at each L. The cross correlation between each explanatory parameter and the MeV electron flux was found to be maximum for the time delay between 0 and 10 days.  As an effort to progress toward more operational forecasting, \cite{2019SpWea..17..438C} proposed a model called PreMeVE to forecast MeV electron variation with lead times of 5 hours, 1 day, and 2 days, for L-shells between 2.8 and 7, based on the linear prediction filters. This model ingest low-Earth orbit (LEO) observation of electron fluxes to forecast the MeV electron based on the cross-energy, cross-L-shell, and cross-pitch angle coherence associated with wave-electron resonant interactions. Furthermore, \cite{2020SpWea..1802399P} developed PreMeVE 2.0 to forecast MeV electron variation up to 2 days in advance using linear regression and artificial neural networks. Here, the LEO observation is taken from the NOAA Polar Operational Environmental Satellite (POES) and the model validation is performed using the Van Allen Probe \cite{2013SSRv..179....3M} and the Los Alamos National Laboratory geosynchronous satellite (LANL-01A). In addition, \cite{2020SpWea..1802402C} proposed the SHELLS (Specifying High-Altitude Electrons Using Low-Altitude LEO Systems) model using artificial neural network to nowcast the outer electron radiation belt at 350 keV and 1 MeV. This model uses only the LEO measurements from POES and the geomagnetic Kp index as inputs, and it is validated using the Van Allen Probe data. 

Inspired by the PreMeVE model series \cite{2019SpWea..17..438C, 2020SpWea..1802399P, 2024SpWea..2203975F}, we consider developing equivalent models for the outer-belt MeV electron flux forecasting but with possible continuous validation. Specifically, as the radiation belt dedicated mission--the Van Allen Probe (launched in August 2012)-- has already been decommissioned since 2019, it would be impossible to continuously validate the aforementioned models. Although the Arase (ERG) mission \cite{2018EP&S...70..101M}, launched in 2016, could be used to validate radiation belt electron variation up to at least 2025, the mission has now been operating beyond its planned lifetime of five years and it will be decommissioned some time in the future. Instead, we consider the POES NOAA satellite series which have been providing continuous monitoring of radiation belt electrons for over 2 solar cycles. Despite the decommissioning of the current POES NOAA-15 (1998--2025), the European equivalent satellite--MetOp-B (launched in 2012)--from EUMETSAT will continue to monitor this environment with the same set of instrument onboard NOAA POES satellites. These long-term datasets thus offer a possibility for continuous validation and updated training useful for model development in the future.   

To model sequential data, recurrent neural networks (RNN) such as Long Short-Term Memory (LSTM) have been the gold standard. In certain cases, temporal convolutional neural networks have also been used. For sequence transduction tasks, recurrent or convolutional neural networks with an encoder and a decoder are typically used. Since the development of the Transformer model \cite{vaswani2017attention}, there has been rapid progress in sequence modelling, particularly in Natural Language Processing (NLP). The development of foundation models — ML or deep learning models trained on vast datasets so that they can be applied across a wide range of use cases — has led to the development of large language models (LLMs) at the core of generative artificial intelligence (AI). Since then, foundation models have been developed in domains other than NLP, including physics \cite{nguyen2025physixfoundationmodelphysics, wiesner2026physicsfoundationmodel}. In particular, Surya — a foundation model for heliophysics — has been developed using decades of solar observations from the Solar Dynamics Observatory, with applications in space weather, such as solar flare forecasting \cite{roy2025suryafoundationmodelheliophysics}. In this work, in addition to classic ML and neural network models, we will also focus on the application of foundation models. Specifically, we will focus on the Time Series Foundation Model (TimesFM), a pre-trained time series model for temporal forecasting \cite{das2024decoderonlyfoundationmodeltimeseries}. To our knowledge, this is the first time that such a model has been used for space weather forecasting.

The paper outline is as follows. We first present the datasets in Section~\ref{sec:data} followed by the methodology including pipeline overview, supervised ML algorithms and TimesFM in Section~\ref{sec:method}. We describe the data preparation, model training, and evaluation metrics in Section~\ref{sec:data-prep-training}. Then, we present the results in Section~\ref{sec:results} focusing on 6-hours forecast in advance and a brief overview on longer forecast horizons up to 3 days, followed by the discussion in Section~\ref{sec:discussions}. Finally, we present the conclusions and perspectives in Section~\ref{sec:conclusions}.


\section{Data} \label{sec:data}

We mainly use electron measurement data from the POES-15 satellite and the GOES satellite series covering data from 2013 to 2024. We also use the solar wind and interplanetary magnetic field (IMF) data measured upstream of Earth from satellites near the Lagrange L1 point, and some geomagnetic indices from ground magnetic measurements. In this Section, we describe in detail the data products and the data processing. 

\subsection{Electron data from POES NOAA-15}

We utilize data from the SEM-2 instrument onboard POES-15 for which the processed level 1b (L1b) data are available from 2012 to 2025 \cite{sem2-user-manual}. Here, we obtain L1b data from the Medium Energy Proton/Electron Detector (MEPED) sensor. The flux measurements from the electron channels E2 ($>$100 keV) and E3 ($>$300 keV), and the proton channel P6 ($>$6.9 MeV) from the 90$^o$ telescopes are used. The 90$^o$ telescopes essentially monitor trapped particles in the Van Allen radiation belts at high latitudes \cite{2008JGRA..11310211G, 2010JGRA..115.4202R}. Although the P6 channel was designed for proton flux measurements, it was found that the counts are mostly contaminated by electrons above 700 keV \cite{2009JASTP..71.1126S, 2010JGRA..115.4202R}. By comparing the P6 measurements to the instrument for the detection of particle  onboard DEMETER satellite, \cite{2019SpWea..17..438C} concludes that the P6 essentially measures $>$1 MeV electrons and thus it can be used quantitatively for the dynamics of precipitating $>$1 MeV electrons.

The L1b data consist of geographical location of POES-15 along its polar orbit, the McIlwain's L-parameter calculated from the International Geomagnetic Reference Field (IGRF) field model, the Magnetic Local Time (MLT), and the counts (in particles cm$^{-2}$ s$^{-1}$ sr$^{-1}$ for E2 and E3, and in particles cm$^{-2}$ s$^{-1}$ sr$^{-1}$ keV$^{-1}$ for P6). They are temporal sequences with 2-second cadence. We obtain all these parameters from January 2013 to May 2024. The original 2s data are averaged to 1h cadence and the L-values are rounded to one decimal place. Only L-values between 2.8 and 6.0 are kept. Then, the data are grouped by time (hourly) and L-values using the mean values. Here, we obtain the tabular output with E2, E3, P6 measurements at each L as a function of time.

\begin{figure}[p]
    \centering
    \includegraphics[width=0.95\linewidth]{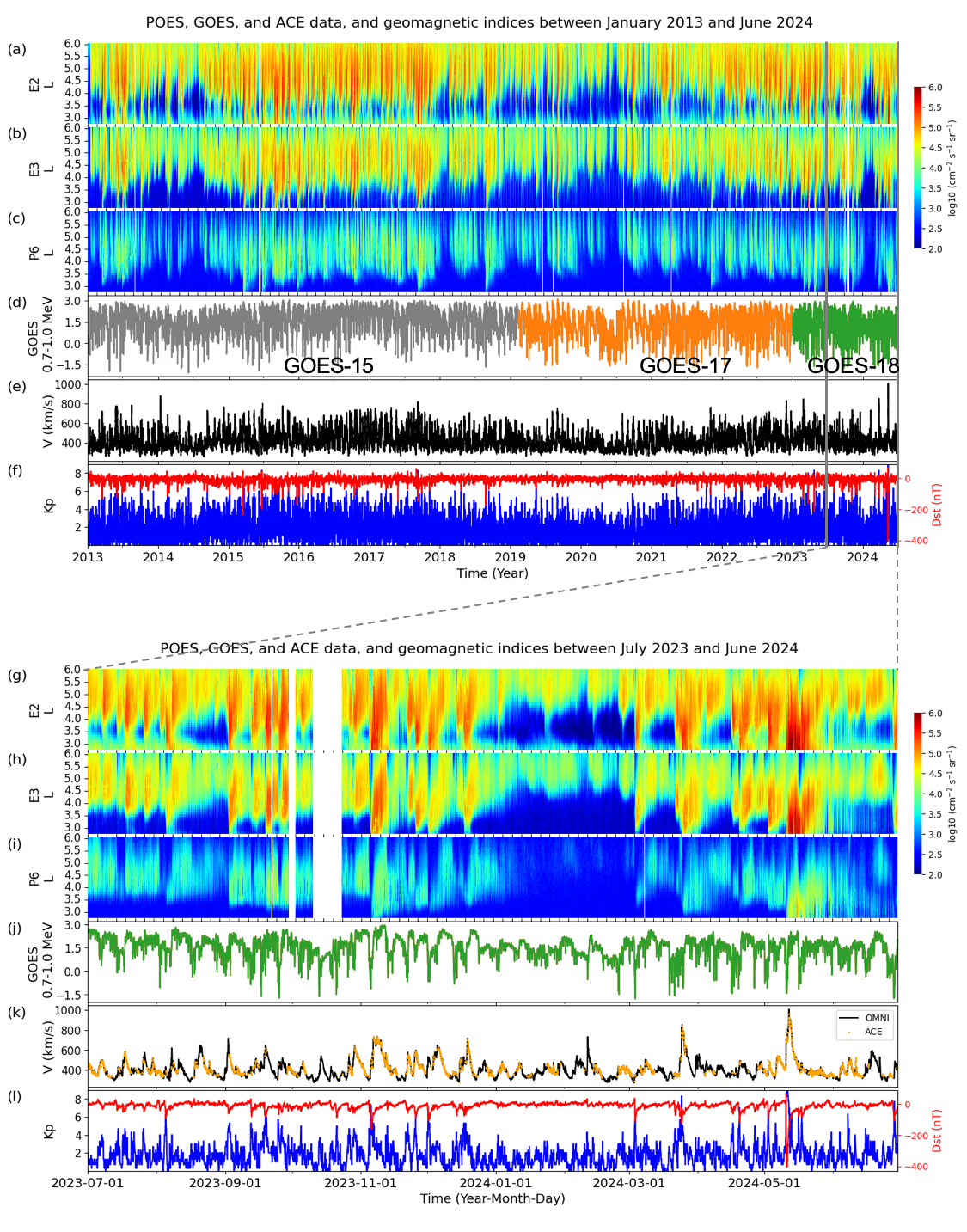}
    \caption{Overview of the data between January 2013 and June 2024 (a--f), and their zoom-in (g--l): (a, b, c) the POES E2, E3, and P6 flux measurements, respectively; (d) the merged GOES data using GOES-15 (grey), GOES-17 (orange), and GOES-18 (green); (e) the solar wind speed from OMNI database; (f) the Kp index (blue) and the Dst index (red); (g--l) same data as in (a--f), respectively, except in (k) where the solar wind speed from ACE (orange) is also shown. }
    \label{fig:data-overview}
\end{figure}

Figures~\ref{fig:data-overview}a, \ref{fig:data-overview}b, and \ref{fig:data-overview}c show the POES E2, E3, and P6 flux measurements, respectively, as a function of L and time, also called ``L-time diagram", for the period between January 2013 and June 2024. The flux values are transformed with the base-10 logarithm. We can see that the E2, E3, and P6 fluxes are more intense with their high values reaching low L values during the solar maxima around 2015--2017 and 2022--2024, compared to during the solar minima around 2014 and 2020. Figures~\ref{fig:data-overview}g, \ref{fig:data-overview}h, \ref{fig:data-overview}i show the zoom-in of the L-time diagrams between July 2023 and June 2024. There are some missing data, seen as white gaps, around October 2023 for instance, which are later excluded from the evaluation. During May 2024, we find the intense E2 and E3 flux enhancements corresponding to the 10 - 12 May 2024 geomagnetic storm with the storm disturbance (Dst) index reaching -400 nT (Figure~\ref{fig:data-overview}l, red line). The P6 measurements, corresponding to the 1-MeV electron flux, also show enhancement at low L, likely reaching the slot region and the inner radiation belt. Detailed observations of this extreme geomagnetic storm with particular effects in the radiation belts can be found in \cite{2024Univ...10..391P}.    

\subsection{Electron data from GOES satellite series} \label{subsec:GOES-data}

We utilize data from the Energetic Proton, Electron and Alpha Detector (EPEAD) instruments from GOES-15 which are available from 2011 to 2020 (Figure~\ref{fig:data-overview}). The energy channel E1 for $>$.8 MeV electron flux measurements (in particles cm$^{-2}$ s$^{-1}$ sr$^{-1}$) for both East and West cameras is used. GOES-15 was replaced by GOES-17 (launched in 2018) and then by GOES-18 (launched in 2022) as the operational GOES-West satellite. To produce a long dataset of the GOES electron fluxes, we merge the data from GOES-15, 17, and 18. Here, the $>$.8 MeV electrons from GOES-15 in 2011 - 2020 are merged with the measurements from the electron channel E7 (0.7 to 1.1 MeV) of the Magnetospheric Particle Sensor - High Energy onboard GOES-17 (2019-2023) and GOES-18 (from late 2022 onwards). The L1b data at 5-minute cadence are used. The flux values are averaged over all the cameras. The data are resampled to 1 hour cadence using the average. The averaged 1-hour flux values are transformed with the base-10 logarithm.

To merge the data from the different GOES satellites, we consider the overlapping time interval between GOES-15 and GOES-17 in 2019. Since GOES-17 and GOES-18 carry the same set of the instrument, the electron measurements can directly be merged by replacing GOES-17 with GOES-18 directly from January 4, 2023 (its operational date), onward. Considering the data in 2019, we find that GOES-15 (E1) and GOES-17 (E7) have the same distributions but with an offset between their centers. In order to scale the GOES-15 data, we perform a linear fitting in the log10 space between GOES-15 and 17 for the data in 2019. The linear fitting function is found to be $y_{15, S} = 1.01 y_{15, O} - 2.46$, where $y_{15, O}$ is the original GOES-15 E1 flux in log10 and $y_{15, S}$ is the scaled GOES-15 E1 flux in log10. The cross-correlation between the scaled GOES-15 E1 flux and the GOES-17 E7 flux in 2019 is found to be 0.97. We perform this scaling to all GOES-15 data since January 1, 2011. The scaled GOES-15 data are replaced by the original GOES-17 data from February 12, 2019 (its operational date) onward. 

With the above procedure, we obtain a long, consistent data set of the near-$1$ MeV electrons from GOES. Figure~\ref{fig:data-overview}d shows this merged data between January 2013 and June 2024 with the scaled GOES-15 (gray), and the original GOES-17 (orange) and GOES-18 (green). Figure~\ref{fig:data-overview}j shows the zoom-in between July 2023 and June 2024.  

\subsection{Exogenous parameters: solar and geomagnetic activities} \label{subsec:sw-geo}

The solar wind and interplanetary magnetic field (IMF) conditions upstream of Earth are crucial for modeling the dynamics of the  magnetosphere and the Earth's radiation belts. We obtain these parameters from the OMNI hourly data set \cite{Papitashvili_King_2020} that is the hourly averaged, multi-source, near-Earth solar wind magnetic field, plasma and energetic proton flux data, plus geomagnetic indices. Here, the satellite data near L1 (ACE or Wind) are shifted to the bow shock nose; the data are cleaned and they are the most complete compared to those of L1 satellites. The caveat of using OMNI data is that they are not available in real time, and thus could pose a problem when transitioning to operational forecasting. For this reason, we also considered using data from ACE. Figure~\ref{fig:data-overview}k shows a comparison between the solar wind speed from OMNI (black) and from ACE (orange), for the Level-2 data, between July 2023 and June 2024. We find that ACE contains a lot of bad values (noted as -1e+31 in the data) that cannot be used for our modeling. While using real-time data, e.g., from ACE, would be desirable, we decided to use OMNI data instead as they contain more complete data. In brief, we obtain the proton density ($N$), the proton speed ($V$), and the IMF $B_z$ (north-south) component, from the OMNI data set. Additionally, we compute the dynamic pressure of the solar wind: $P_{dyn} = \rho V^2 / 2 $, where $\rho = N m_p$ is the proton mass density and $m_p$ is the proton mass.

In addition to the solar wind and IMF conditions, we also obtain the geomagnetic indices from ground magnetic measurements. In particular, we obtain the hourly storm disturbance index (Dst) \cite{sugiura1963hourly} from the World Data Center for Geomagnetism, Kyoto \cite{1460283805892641024}. We also obtain the 3-hour cadence Kp index \cite{bartels1949standardized} from the GFZ Helmholtz Centre for Geosciences \cite{matzka2021geomagnetic}. These geomagnetic indices are obtained for January 2013 - June 2024 as shown in Figures~\ref{fig:data-overview}f and \ref{fig:data-overview}l. 


\section{Methodology} \label{sec:method}

We first describe the pipeline overview including model inputs and output in Section~\ref{subsec:pipeline}. It is followed by the model overviews focusing on the ML algorithms including the linear regression, neural networks, Transformer encoder in Section~\ref{subsec:ml-archi-inputs}, and the time series foundation model in Section~\ref{subsec:timesfm}. We describe in particular how each ML algorithm is used regarding its architecture as well as the input and output data handling. 

\subsection{Pipeline overview} \label{subsec:pipeline}

\begin{figure}[h]
\centering
\includegraphics[width=0.85\textwidth]{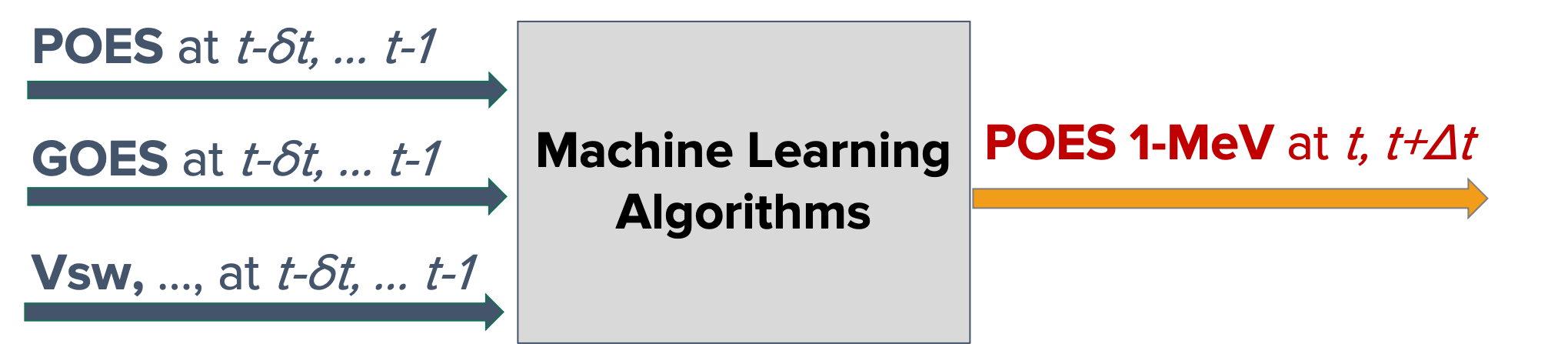}
\caption{Schematic diagram of our pipeline for predicting 1-MeV electrons in the outer radiation belt using past sequences of precipitating electrons in the low-Earth orbit from POES, near 1-MeV electrons from GOES, and exogenous parameters including the solar and geomagnetic activity indicators.}
\label{fig:schematic-pipeline}
\end{figure}

We exploit the relationship between the precipitating low energy electrons ($>$100 and $>$300 keV) and the relativistic ($>$1 MeV) electrons in the outer radiation belts. Following the development of PreMevE-2.0 \cite{2020SpWea..1802399P}, we adopt the same philosophy of predicting 1 MeV electrons at each L-value. The L-shell domain is defined from L=2.8 to L=6.0 with $0.1L$ resolution. Individual models are developed for the L values (with a total of 33 L-values), meaning that these models are optimized separately using different sets of input variables. For all L-values, the cross L-shell information on the E2 flux measurement at $L=4.6$ is used ; it was found by \cite{2019SpWea..17..438C} that the intensified precipitation of $>$100 keV electrons at $L = 4.6$ always leads MeV electron energization. Later, the prediction results will be shown for each L or concatenated together to produce a continuous domain of the outer radiation belt ($L \in [2.8, 6.0]$). Figure~\ref{fig:schematic-pipeline} shows a schematic of the pipeline for 1-MeV electron prediction using the past sequences of POES and GOES electrons and exogenous variables as inputs. 

For modeling $1$-MeV electron flux at each L, we use past sequences of the following data as inputs. First, the past time series of POES E2, E3, and P6 flux values and the magnetic local time (MLT) at the corresponding L are used. Second, the merged GOES electron flux is used. Third, the exogenous variables ($V$, $N$, $P_{dyn}$, $B_z$, Dst, Kp) are used. Figure~\ref{fig:schematic-pipeline} summarizes the main inputs in grey (left arrows). In addition, the temporal derivatives of E2 fluxes, namely dE2, defined as $dE2_{t} = (E2_t - E2_{t-1}) / E2_{t-1}$ at each L, is also used as inputs to incorporate information on the onset timings of MeV electron events \cite{2019SpWea..17..438C, 2020SpWea..1802399P}. For all L values, the E2 flux at L=4.6 from POES and the near 1-MeV electron flux from GOES are used as the cross L-shell information. This set of input variables are identical to all the following models (except for the TimesFM, which will be explained in Section~\ref{subsec:timesfm}). 

\subsection{Supervised machine learning algorithms} \label{subsec:ml-archi-inputs}

Here we select a subset of models presented in \cite{2020SpWea..1802399P}: linear regression, 1D convolutional neural network (Conv1D), and long short-term memory neural network (LSTM). Additionally, we implement a Transformer-Encoder (TransEncode) model that includes only the encoder part of the Transformer \cite{vaswani2017attention}. Specifically, these models are used for multi-variate time series regression with multiple variables (past sequence) as inputs and a single output (1-MeV electron flux). 

\subsubsection{Linear regression}

Linear regression is a simple method for fitting linear relationship between input and output variables. To take into account the past variation, the input variables are shifted manually in time from $t-1$ to $t-32$. Considering the 6-hour cadence data, the shifted data include past variation up to 8 days ($6 \times 32 = 192$ hours). For each L, we have 12 input variables. Shifting them from $t-1$ to $t-32$ yields $12 \times 32 = 384$ input sequences for the prediction at each L. Here we perform multi-variate linear regression for the target P6 at $t, t+1, ...$ at each L, each of which takes 384 input variables. 

\subsubsection{1D convolutional neural network (Conv1D)}

Conv1D is a variant of the CNN adapted for temporal sequences \cite{kiranyaz20191dconvolutionalneuralnetworks}. Unlike the classic CNN, the Conv1D uses convolutional filters only in one dimension. This allows the nodes to extract important local characteristics such as the temporal dependence or pattern on the inputs. Similar to above, we consider the past sequences up to $t-32$. This defines the length of the window of the inputs (i.e., 1D image of size 32) to be fed to the network to model our target at each L. The input layer consists of one Conv1D layer. The Rectified Linear Unit \cite{glorot2011proceedings} is used as the activation function to introduce nonlinearity. A dropout layer \cite{srivastava2014dropout} with a dropout ratio of 0.2 is added before the output layer to prevent overfitting. 

\subsubsection{Long short-term memory neural network (LSTM)}

LSTM is a variant of the RNN specifically designed for processing and predicting temporal sequences \cite{Hochreiter1997LongSM}. Unlike RNN, the LSTM uses a memory cell consisting of gates for input, output, and forgotten information that allow them to efficiently conserve long-term temporal dependence while incorporating short-term information into the forecast. Here, the input sequences of size 32 is used. The network consists of an LSTM layer with hidden size of 128 followed by a dropout layer (dropout rate = 20\%). Our LSTM network is trained separately for the prediction at each L.  

\subsubsection{Transformer Encoder (TransEncode)}

TransEncode has the Transformer architecture \cite{vaswani2017attention} but consists of only the encoder part, the multi-head attention, and feed-forward neural network (for the output). The multi-head attention is the parallelized self-attention mechanism \cite{kim2017structured} that efficiently captures the context (i.e., temporal dependency) in long sequences. Our TransEncode network has two encoder layers to inject positional information into the input sequences (of length 32 similar to above). The transformer regressor itself has the embedding dimension of 64 and two attention heads. Finally, the feed-forward layer has the dimension of 256 and the dropout rate of 0.1 is added. 

\subsection{Time series foundation model (TimesFM)} \label{subsec:timesfm}

TimesFM is a pre-trained time series foundation model developed by Google Research for time-series forecasting \cite{das2024decoderonlyfoundationmodeltimeseries}. It is a decoder-style attention model with input patching, similar to a token in language models. It is designed for univariate time-series forecasting (auto-regressive modeling) without further training for context lengths (i.e., past values) up to 2048 time steps and any horizon lengths (i.e., forecast time steps). It has been trained using a large time-series corpus with different time cadences from hourly to monthly. For our specific application, we use TimesFM 2.0 (timesfm-2.0-500m-pytorch) that has been trained on a subset of the Large-scale Open Time Series Archive (LOTSA) \cite{woo2024unifiedtraininguniversaltime}. This subset of data includes time series in diverse sectors including the energy, transport, climate, nature, web, economy, finance, and cloud operation services. We note that there is no space weather data in this pre-training data subset, although the sunspot number is included in the full LOTSA dataset. 

We apply TimesFM in two distinct ways : (1) we first elaborate the default application of TimesFM, namely the TimesFM zero-shot in Section~\ref{subsubsec:tfm-zero}, and (2) the hybrid application of TimesFM, namely the TimesFM with covariates in Section~\ref{subsubsec:tfm-cov}. 

\subsubsection{TimesFM zero-shot} \label{subsubsec:tfm-zero}

TimesFM was designed to be a general purpose zero-shot forecaster: it takes in the past $C$ time-steps of a time series as ``context" and predicts the future $H$, i.e., horizon, time-steps. Given the context denoted by $\mathbf{y}_{1:L} := \{y_1, ..., y_L\}$, we aim to predict values in the horizon $\mathbf{y}_{L+1:L+H}$. As it was built for a general purpose forecasting, it cannot take other variables, i.e., static or dynamic covariates, as context by default. For instance, if we were to predict a local temperature for the next $n$-steps, we can only take the past local temperature measurements over $C$ points but we cannot take other measurements such as the humidity, luminosity, as context (i.e., inputs). In brief, the TimesFM zero-shot (subscript: $tfm$) can be used to map any time series context to horizon,
\begin{equation}
f_{tfm} : (\mathbf{y}_{1:L}) \rightarrow \hat{\mathbf{y}}_{L+1:L+H}
\end{equation} 

To predict the 1-MeV (P6) electron flux in the outer radiation belt, the only input variable is therefore the past values of P6 flux itself. Similar to above, our horizon length $H$ is set to be 1 step (6 hours for the 6h cadence data). For the context length $C$, we first attempted with 32 time steps as for the other ML models in Section~\ref{subsec:ml-archi-inputs}. Since this took a rather long computational time, we experimented varying the context length from 1 to 32 time steps, i.e., from 6 hours to 8 days. The experiment was conducted for the 1-MeV electron flux prediction at $L=4.0$ for the data between July and December 2023. It was found that the $R^2$ score no longer increases beyond the context length of 8 time steps. Therefore, we fixed the context length $C$ to be 8 time steps for the TimesFM.   

Unlike the ML algorithms above, we note that the TimesFM is not re-trained by our dataset. In the next Section, however, the exogenous variables in Section~\ref{sec:data} for the time range equivalent to the neural network model validation is used for fine-tuning.

\subsubsection{TimesFM with covariates (TimesFM+Cov)} \label{subsubsec:tfm-cov}

Our problem is indeed formulated as a multi-variate forecasting where the 1-MeV electron flux can be predicted using past variations of the precipitating electrons, solar wind and IMF conditions, and geomagnetic indices. To incorporate exogenous variables in modeling with TimesFM, the covariates, i.e., the variables that co-vary with the target, are introduced in Appendix A1 of \cite{das2024decoderonlyfoundationmodeltimeseries}. In principle, we can have two sub-models: the first model is the linear regression (``xreg") of the time series itself on the covariates, and the second model is the forecast of the residuals using TimesFM. The final results are the sum of the prediction from these two sub-models (``xreg + timesfm"). 

Generally, covariates can be categorized as ``static" and ``dynamic" covariates; the former changes with time and the latter do not. A key limitation of the described approach is that the dynamic covariates need to cover both context and horizon. However, the variation of dynamic covariates cannot be known in advance (i.e., covering the horizon time) without predicting them, which would result in cumulative prediction errors. To overcome this, it was suggested\footnote{see \url{https://github.com/google-research/timesfm/blob/master/v1/notebooks/covariates.ipynb}} that the past dynamic covariates that are only available for the context could be used. Concretely, the covariates could be shifted and repeated to use their delayed version, also called the past dynamic covariates. In other words, we can use the past variations of the exogenous variables as inputs to the first sub-model. 

To apply TimesFM with covariates in our case, the first step consists in ridge regression between the past dynamic covariates and the target (P6), and the second step consists in applying TimesFM zero-shot on the residuals from the ridge regression. Ridge regression is similar to the ordinary least squares regression but with an added penalty term; it seeks to minimize the following: 
\begin{equation}
\textrm{RSS} + \lambda \sum \beta_j^2 \label{eq:rss-ridge}
\end{equation}
, where $\textrm{RSS} = \sum (\mathbf{y}_i - \hat{\mathbf{y}}_i)^2$ is the sum of squared residuals, $\mathbf{y}_i$ is the actual value of the $i^{th}$ observation and $\hat{\mathbf{y}}_i$ is the predicted value for the $i^{th}$ observation, $\beta_j$ are the coefficients in multiple linear regression, i.e., $Y = \beta_0 + \beta_1 X_1 + \beta_2 X_2 + ... $. The second term in (\ref{eq:rss-ridge}) is known as a shrinkage penalty. When $\lambda=0$, this penalty term has no effect and ridge regression produces the same coefficient estimates as least squares. Here, ridge regression is chosen over linear regression as the former can reduces the effects of multicollinearity where two or more input variables are highly correlated with each other. After applying ridge regression, we can obtain the residuals of the target for the context:
\begin{equation}
\mathbf{y}_{res, i:L} = \mathbf{y}_{i:L} - \hat{\mathbf{y}}_{reg, i:L} \label{eq:res-comp}
\end{equation}
, where $\mathbf{y}_{res, i:L}$ is the residuals covering the context ($i:L$), $\mathbf{y}_{i:L}$ is the observed value for the context, and $\hat{\mathbf{y}}_{reg, i:L}$ is the predicted value from ridge regression, i.e., 
\begin{equation}
f_{ridge}: (\mathbf{x}_{i:L}) \rightarrow \hat{\mathbf{y}}_{reg, i:L} \label{eq:ridge-reg}    
\end{equation}
, where $\mathbf{x}_{i:L}$ are the past dynamic covariates. Then, TimesFM is applied on the residuals: 
\begin{equation}
f_{tfm} : (\mathbf{y}_{res, 1:L}) \rightarrow \hat{\mathbf{y}}_{res, L+1:L+H} \label{eq:res-proj}
\end{equation} 
, where $\hat{\mathbf{y}}_{res, L+1:L+H}$ is the predicted residuals covering the horizon ($L+1:L+H$). As mentioned, $\hat{\mathbf{y}}_{reg, i:L}$ is computed from the past dynamic covariables. Since the dynamic covariables must cover both context and horizon, and that we shifted the dynamic covariables to use their delayed version, we can indeed obtain the predicted $\hat{\mathbf{y}}_{reg}$ by shifting back the time labels: 
\begin{equation}
\hat{\mathbf{y}}_{reg, i:L} \rightarrow \hat{\mathbf{y}}_{reg, L+1:L+H}. \label{eq:reg-exo}
\end{equation}
Finally, we can obtain the final prediction from the sum of the ridge regression and residual prediction:
\begin{equation}
\hat{\mathbf{y}}_{L+1:L+H} =  \hat{\mathbf{y}}_{reg, L+1:L+H} + \hat{\mathbf{y}}_{res, L+1:L+H}. \label{eq:tfm-cov-final} 
\end{equation}

Similar to the classic models in Section~\ref{subsec:ml-archi-inputs}, we use the shifted exogeneous variables ($V$, $N$, $P_{dyn}$, $B_z$, Dst, Kp) as past dynamic covariates. However, for the variables regarding the precipitation electrons, our TimesFM with covariates application takes slightly different sets of inputs. Here, the shifted sequences of POES E2, E3, dE2, and P6 fluxes at all L values except the P6 flux at the target L, and the shifted electron flux from GOES, are used as past dynamic covariates. This choice is motivated by the definition of covariates themselves that they must vary (but do not coincide) with the target.


\section{Data preparation and model training} \label{sec:data-prep-training}

After describing the data sources and our ML algorithms, here we further detail the data processing to be ML ready and the training process before reporting the results in Section~\ref{sec:results}. The data preparation and scaling are described in Section~\ref{subsec:data-scaling}. The details on model training are given in Section~\ref{subsec:mod-training}. In particular, one problem when training a model for time series prediction is the data bias owing to time-dependency. We address this point in Section~\ref{subsubsec:walk-forward} used specifically for the neural network training in Section~\ref{sec:network-training}.

\subsection{Data preparation and scaling} \label{subsec:data-scaling}

To facilitate the data ingestion into machine learning training, all time series data in Section~\ref{sec:data}, e.g., in Figure~\ref{fig:data-overview}, are concatenated to a single tabular data. This dataset is then resampled using average to 6h, 12h, and 24h time resolutions. The null values are excluded and the linear interpolation is performed to fill nulls when presented. 

To scale the data, each variable is scaled separately using the robust scaler: 
\begin{equation}
x_{scaled} = \frac{x - x_{median}}{x_{q3} - x_{q1}}, \label{eq:scaler}
\end{equation} 
where $x$ is the original value, $x_{median}$ is the median, $x_{q3}$ is the 75th percentile and $x_{q1}$ is the 25th percentile of the data. The median, q1, and q3, are computed for the data range between January 2013 and December 2023, which is taken as the training data for the neural networks. The scaling parameters are saved individually for each variable and then applied to all data range (January 2013 - May 2024). We note that the robust scaler is chosen instead of the min-max scaler or standard scaler because it is more robust to outliers and is well adapted to data with skewed distribution. 

To further facilitate the data ingestion for the model training later, we save our scaled data into two time tables: one with all parameters except the targets, and one with only the target variables (POES P6 channel) from January 2013 to May 2024. 

\subsection{Model training} \label{subsec:mod-training}

Our machine learning algorithms can be categorized into three groups: (1) linear regression, (2) neural network models, and (3) TimesFM. As the architectures and designs of these algorithms are different, we adopt different strategies for model fitting for them. For TimesFM, there is no need for training as the model is already pre-trained. We perform nevertheless a fitting on the ridge regression when using the TimesFM with covariates as introduced in Section~\ref{subsubsec:tfm-cov}. For the neural network models, we train and validate them using a large dataset between January 2013 and December 2023 and test them in January - May 2024; this is explained in Section~\ref{sec:network-training}. In particular, we adopt the walk forward validation for these networks explained in Section~\ref{subsubsec:walk-forward}. For both linear regression and ridge regression in TimesFM with covariates, we use data in June - December 2023 for the model fitting to avoid an over-generalization of the models. Similar to the neural networks, we use data in January - May 2024 for the test.

As mentioned in Section~\ref{subsec:ml-archi-inputs}, we train our models separately for predicting 1-MeV electron flux at each L-shell. In other words, for each ML algorithm, there are 33 sub-models for $L=2.8, 2.9, ... 6.0$. These sub-models are trained independently from each other although with some common inputs taken as cross L-shell information. 

\subsubsection{Neural network training} \label{sec:network-training}

Here we focus on Conv1D, LSTM, and TransEncode models described in Section~\ref{subsec:ml-archi-inputs}. The training process is aimed at searching for optimum weights and biases that yield minimum loss using the training data for each model, until it reaches satisfying performance on the validation data. For all these models, the weights and biases in the neural networks are initialized randomly with a fixed seed number to ensure reproducible results. The models are set up to learn from a certain amount of data, i.e., learning in batches, with the batch size = 1024 at each learning cycle. Through the learning cycles known as ``epochs", these weights and biases are gradually optimized. The optimization of learning is done through the stochastic gradient descent using the adaptive moment estimation, known as ``Adam" algorithm \cite{kingma2014adam}, with the learning rate of 0.001. At the end of each learning cycle (epoch), the loss function, defined to be the Mean Square Error (MSE) is computed for both training and validation data. The learning is stopped when there is no improvement on the validation loss for 10 consecutive steps. We note that the maximum epoch is set to 100 for all models. The best weights and biases are saved at the best epoch where the validation loss reaches its minimum. 

\subsubsection{Walk-forward Validation} \label{subsubsec:walk-forward}

A common problem when training a model with large datasets is the partitioning between training, validation, and test sets. Depending on the partitioning, it can introduce biases such that the training data may not have the same distribution nor representative enough for the validation and test data. Unlike time-independent problems, we cannot shuffle temporal sequences. Therefore, we need a technique for data splitting for model training that allows us to split data while keeping temporal order of observations. In particular, the space weather data are strongly influenced by the solar cycle where the Sun becomes more active (reaching solar maximum) with its surface undergoes more frequent eruptions and then becomes more calm (reaching solar minimum) over 11-year cycle. During solar maximum where there are more occurrences of solar-driven transients (e.g., ICMEs and high-speed streams) hitting the Earth, the energetic electron fluxes in the outer radiation belt become highly varying, sometimes reaching low L-shells, following the electron precipitation and various processes in the magnetosphere \cite{MIYOSHI201177}. In the absence of solar-driven transients during solar minimum, however, the energetic electron fluxes in the outer radiation belt become absent or slowly varying \cite{LAZUTIN20172248}. To account for this temporal dependency influenced by the solar cycle, we adopt a technique called ``Walk-Forward Validation" as described next. 

\begin{figure}[h]
\centering
\includegraphics[width=0.95\linewidth]{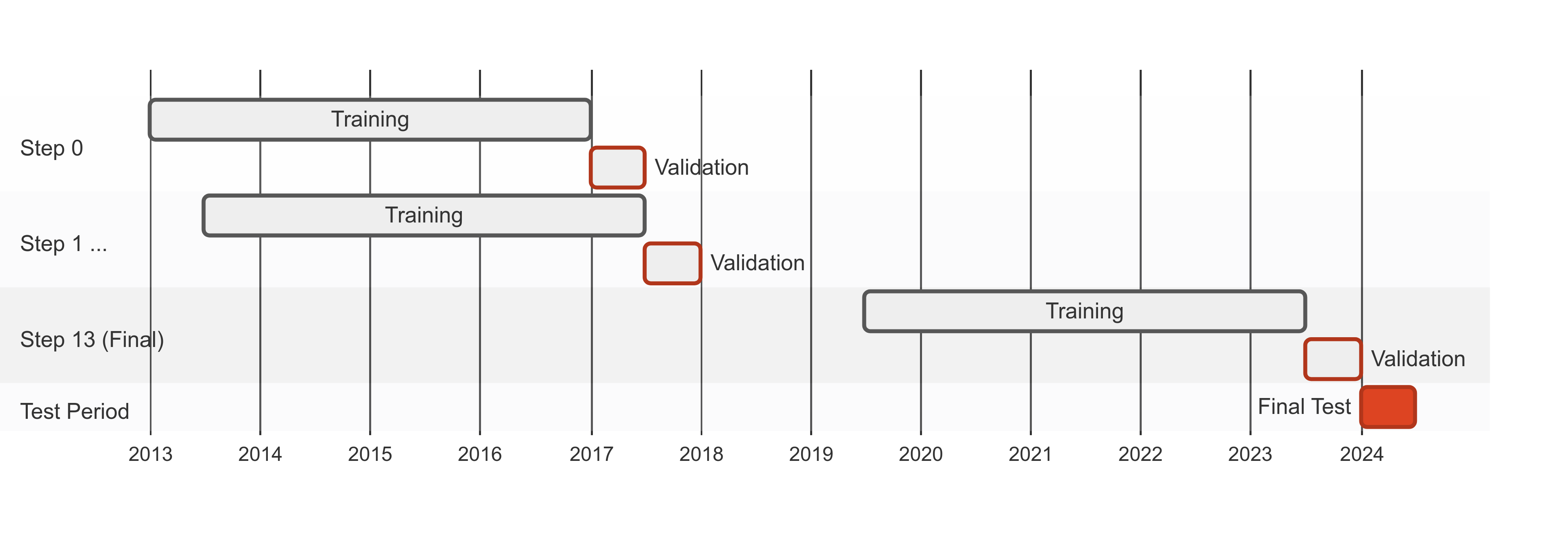}
\caption{Schematic illustration of the walk-forward validation.}
\label{fig:walk-forward}
\end{figure}

The walk-forward validation is a technique commonly used for financial time series and weather forecasting. It allows us to retrain our model once new data become available. For this reason, it is rather practical for future model updates. The method consists in training and validating a model in several ``walk-forward" steps, as illustrated in Figure~\ref{fig:walk-forward}. Firstly, an initial model is trained from scratch using the data in a defined time window; it is then validated using the (new) adjacent data in a defined validation window. For instance, here we define our training window length to be 4 years and the validation window length to be 6 months (see Step 0 in Figure~\ref{fig:walk-forward}). After training the first model, the model weights and biases (see Section~\ref{sec:network-training}) are saved. Next, the training is resumed from the previous step, i.e., the learned weights and biases are updated, using the training data in the shifted window; then, it is validated in the next, adjacent validation window (Step 1 in Figure~\ref{fig:walk-forward}). These steps are repeated with the sliding (i.e., walk-forward) windows until the end of the data excluding the test data. After all these walk-forward steps, the model is finally tested with out-of-sample data (red block in Figure~\ref{fig:walk-forward}).

The walk-forward validation was employed in space weather forecasting (e.g., geomagnetic baseline prediction) \cite{2025SpWea..2304192K}. Unlike the nested cross-validation, e.g., applied in \cite{2022JGRA..12730868B}, the walk-forward approach allows the model weights and biases to be optimized several times with more-recent data (while forgetting too-old data), making it most relevant to the time closer to the out-of-sample test data. Here, we use a dataset covering 11 years, covering a solar cycle, between 2013 and 2023 to train the neural network models with the walk-forward validation. The data in January - May 2024 is taken as the test data and is excluded from the model training and validation. 

\subsection{Model performance evaluation} \label{subsec:metrics}

In the following, we evaluate the model performance using the $R^2$ score, the coefficient of determination or the goodness of fit of a model, defined as 
\begin{equation}
R^2 = 1 - \frac{\sum_{i=1}^n (y_i - \hat{y}_i)^2}{\sum_{i=1}^n (y_i - \bar{y})^2}, \label{eq:r2-score}
\end{equation}
where $y_i$ is the $i^{th}$ observation (true value), $\hat{y}_i$ is the $i^{th}$ estimated (modeled) value, and $\bar{y}=\frac{1}{n} \sum_{i=1}^n y_i$. It represents the proportion of variance of $\mathbf{y} = [y_0, y_1, ..., y_n] = y_i$ that has been explained by the independent variables in the model. $R^2$ is in range (-$\infty$, 1] where an $R^2$ of 1 indicates that the regression predictions perfectly fit the data; an $R^2$ of 0 indicates that the model always predicts the expected (average) value of $y_i$. When $R^2$ is negative, it means that the model is arbitrarily worse than the climatology model, i.e., using the averaged data as the prediction. This metrics is used in previous similar works such as in \cite{2011SpWea...9.6003T, 2016SpWea..14...22B, 2020SpWea..1802399P, https://doi.org/10.1029/2021SW002822} but is named as the prediction efficiency (PE). As the electron flux can vary by orders of magnitude, we emphasize that the $R^2$ score here is calculated using the log of the observed and modeled values \cite{2016SpWea..14...22B}. 

Furthermore, to quantitatively compare the model with the observation, we also employ the prediction error, computed point by point on the data, defined as 
\begin{equation}
    \textrm{Prediction error} = \frac{\hat{y}_i - y_i}{y_i} * 100 \%. \label{eq:pred-error}
\end{equation}
The prediction efficiency varies between -100 and 100; the positive (negative) value indicates that the model overestimates (underestimates) the observation. This metric is used specifically when comparing the L-time diagram results in Section~\ref{sec:results}. 


\section{Results} \label{sec:results}

We now focus on the results from all the models on the test data between January and June 2024. We first discuss the quantitative and qualitative results for the 6h (one step) forecast in Section~\ref{subsec:results-6h}, then followed by quantitative results on the long-horizon (multi-steps) forecasts up to 72h for our best model in Section~\ref{subsec:results-timesfm}. 

\subsection{Model evaluation for the 6h forecast horizon (1-step forecasting)} \label{subsec:results-6h}

\begin{figure}[h]
\centering
\includegraphics[width=0.95\textwidth]{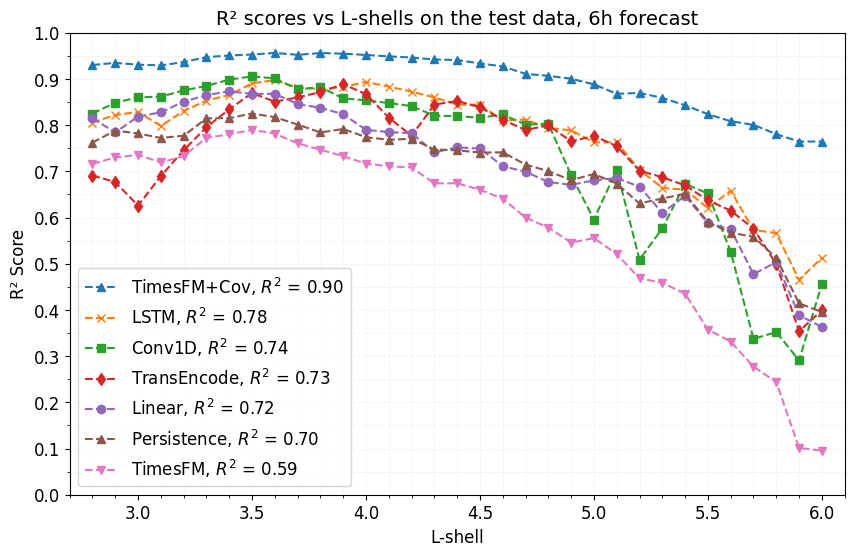}
\caption{$R^2$ scores of the test data (January - May 2024) as a function of L-shells of all the models for 6 hours forecast horizon using 6 hours data cadence.}
\label{fig:r2_lshells_6h}
\end{figure}

Figure~\ref{fig:r2_lshells_6h} shows the averaged $R^2$ score for the 6h horizon forecast over all the test data as a function of the L-shell for the individual models. In addition to those described in Section~\ref{subsec:ml-archi-inputs}, we also add the persistence baseline where the previous 6h observation is used as the next 6h prediction. For all the models, we find that the maximum $R^2$ scores are found near L=3.5, similar to \cite{2020SpWea..1802399P}. These $R^2$ scores decrease for the higher L-shells with the minimum $R^2$ values at L=6.0. When comparing between the models, we find that the TimesFM+Cov (blue triangles) shows the best average $R^2$ score of 0.90, surpassing all the models. The best $R^2$ score of 0.96 of the TimesFM+Cov is found at L=3.6. The second best model is found to be the LSTM (orange crosses) with the average $R^2$ score of 0.78, and the best $R^2$ score of 0.89 at L=3.6. The Conv1D (green squares), the TransEncode (red diamonds), and the linear regression (purple dots) models yield similar average $R^2$ scores of 0.74, 0.73, and 0.72, respectively. These all models perform better than the persistence baseline (brown triangles) with the average $R^2$ score of 0.70. The TimesFM zero-shot (pink triangles) shows the worst average $R^2$ score of 0.59. This shows that the zero-shot application of TimesFM performs much worse than the persistence baseline and when the past dynamic covariates are incorporated. 

In Figure~\ref{fig:r2_lshells_6h}, we find that the TimesFM+Cov outperforms all the models for all L-shells. At the lowest L-shell of 2.8, the second best model is found to be the Conv1D with an $R^2$ of 0.83. The TimesFM+Cov at this L-shell yields an $R^2$ of 0.93, corresponding to about 12\% of the improvement. At the highest L-shell of 6.0, the second best model is found to be the LSTM with an $R^2$ of 0.52. The TimesFM+Cov at this L-shell yields an $R^2$ of 0.77, corresponding to about 48\% of the improvement. Compared to the neural networks which are trained from scratch, the TimesFM+Cov performs much better with significant improvement. On average, the TimesFM+Cov yields 15\% improvement in the $R^2$ score compared to the LSTM model for the L-shells in the range of [2.8, 6.0]. 

\begin{figure}[p]
    \centering
    \includegraphics[width=0.95\linewidth]{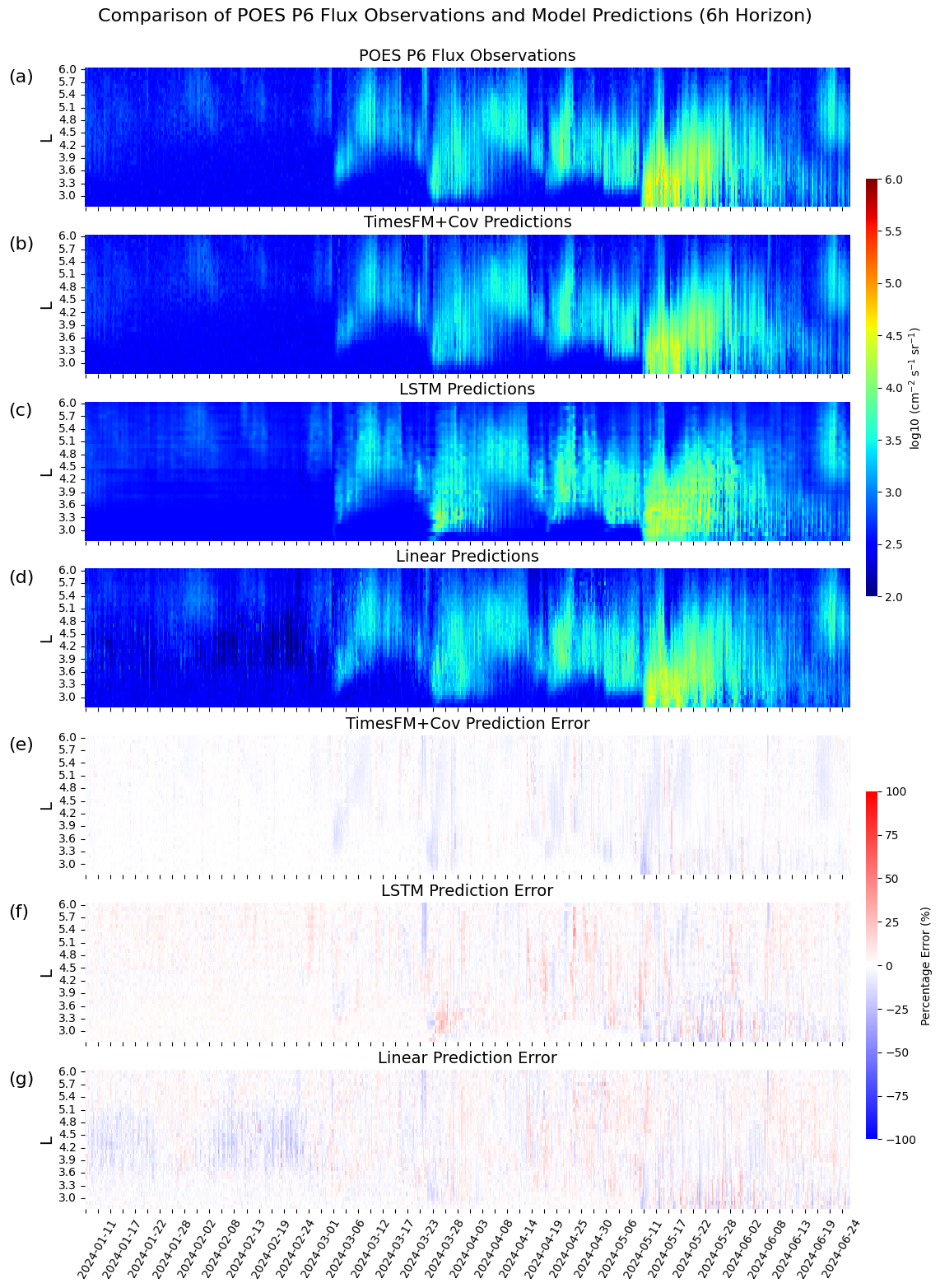}
    \caption{L-time diagrams (a-d) and the prediction error (e-g): (a) the 1-MeV electron flux observations; (b - d) 6h forecast prediction from the TimesFM+Cov, LSTM, and linear models, respectively; (e - g) prediction error of the TimesFM, LSTM, and linear models, respectively.}
    \label{fig:ltime-pred-error}
\end{figure}

Next, we focus on the qualitative results from the TimesFM+Cov, LSTM and linear regression models. Figure~\ref{fig:ltime-pred-error} shows the L-time diagrams of the 1-MeV electron flux observations in panel (a), L-time diagrams of the modeling results from TimesFM+Cov in panel (b), from the LSTM in panel (c), and the linear regression in panel (d). Compared to the observations, we find that all of these models produce similar large-scale variations especially for the May 2024 geomagnetic storm. On the 10th of May 2024, Figure~\ref{fig:ltime-pred-error}a shows the high 1-MeV electron flux seen as green-yellow patches penetrating down to the low L-shells, likely lower than L=2.8. According to \cite{2024Univ...10..391P} and \cite{2026JGRA..13134419S}, the MeV electrons were observed to be injected to the inner belt down to L=2.4, lower than the so-called impenetrable barrier of such relativistic electrons previously estimated to be at L= 2.8 during the Van Allen Probes' era. 

All of the models in Figures~\ref{fig:ltime-pred-error}b, \ref{fig:ltime-pred-error}c, and \ref{fig:ltime-pred-error}d show similar storm onset on the 10th of May. The storm variation appears as the high 1-MeV electron flux in the L-shells between 2.8 and 5.0 for over two weeks up to around the 28th of May. Similar large-scale features namely the storm onset and the storm variation can be seen in all other events prior to May 2024 despite some different details. The LSTM model, for instance, shows rather pixel-like or patchy-like features throughout. Furthermore, during the quiet times characterized by the low 1-MeV flux prior to March 2024, the various models show different details with some visible underestimation or overestimation. 

To qualitatively characterize the differences between the observation and the modeling results, we now focus on the prediction error as defined in Section~\ref{subsec:metrics}. Figures~\ref{fig:ltime-pred-error}e, \ref{fig:ltime-pred-error}f, and \ref{fig:ltime-pred-error}g show the prediction error from the TimesFM+Cov, the LSTM, and the linear regression, respectively. The TimesFM+Cov shows mostly a white area during the non-storm times prior to March 2024, indicating near 0\% of the prediction error. During the storm onset times, however, we observe the blue areas spanning over most L-shells especially on the 10th of May. This shows that the TimesFM+Cov underestimates the 1-MeV electron flux during the storm commencement. This underestimation is also visible around the 20th of March. For the LSTM model, we find mostly red areas, indicating an overestimation of the 1-MeV electron flux, although some underestimation is seen also during the storm times at low L values for the May 2024 storm. For the linear model, we find an underestimation of the 1-MeV electron flux during the non-storm times before March 2024, and some overestimation afterward. In brief, the TimesFM+Cov seems to yield the least prediction error among these models, although some underestimation of the 1-MeV electron flux is usually seen during the storm onsets.

\begin{figure}[h]
    \centering
    \includegraphics[width=0.95\linewidth]{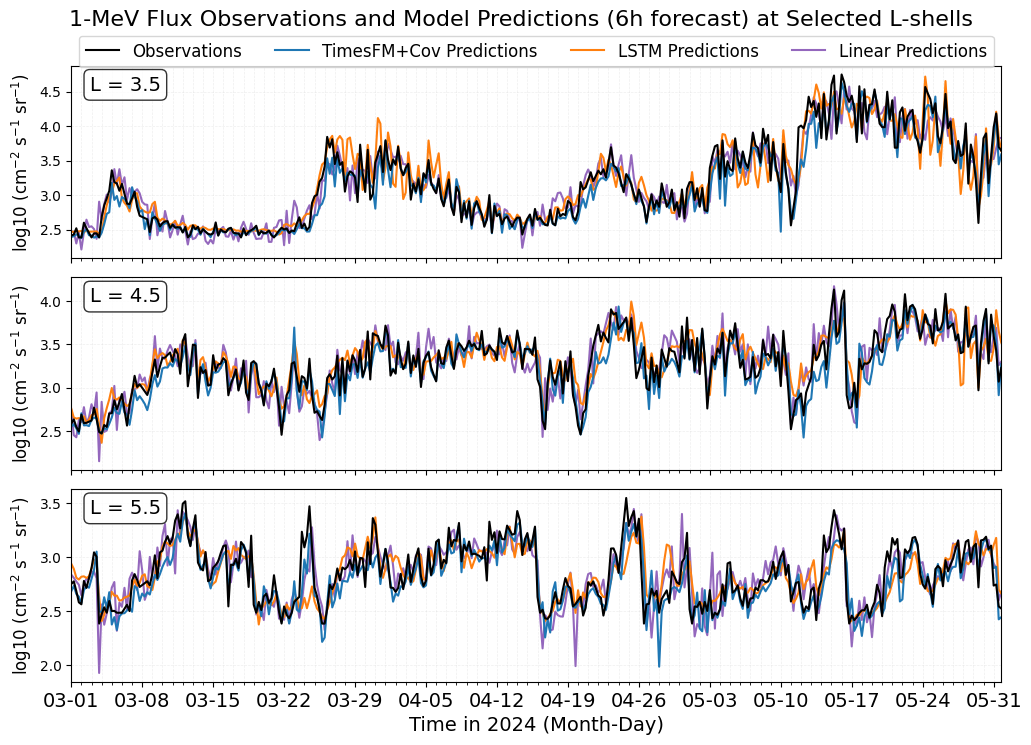}
    \caption{Comparison between the 1-MeV flux observations (black) and the 6h forecast predictions from TimesFM+CoV (blue), LSTM (orange), and Linear (purple) models. The results are shown at (a) L = 3.5, (b) L = 4.5, and (c) L = 5.5 between March and May 2024.}
    \label{fig:obs-pred-selected}
\end{figure}

To better see the differences between the models during the storm times, we now focus on the prediction at certain L-shells. Figure~\ref{fig:obs-pred-selected} shows the comparison between the observations (black) and the modeling results between March and May 2024 from the TimesFM+Cov (blue), the LSTM (orange), and the linear regression (purple). The 1-MeV electron fluxes at L = 3.5, 4.5, and 5.5 are shown in panels (a), (b), and (c) respectively. The storm commencements are characterized by the enhancement in the observed 1-MeV electron flux at low L-shells and the dropout of the observed 1-MeV electron flux at high L-shells \cite{https://doi-org.insu.bib.cnrs.fr/10.1002/2015GL064747, https://doi-org.insu.bib.cnrs.fr/10.1029/2020JA028487}. This pattern can be seen in at L = 3.5 and at L = 5.5 around March 3, March 24, April 19, May 2, and May 10, 2024, corresponding to the major geomagnetic storms with Kp = 6+, 8+, 7, 7-, and 9-, respectively (see also Figure~\ref{fig:data-overview}l). During these storm commencements, we find that all the models produce similar enhancements (Figure~\ref{fig:obs-pred-selected}a) and decreases (Figure~\ref{fig:obs-pred-selected}c) of the 1-MeV electron flux. However, we find that they are slightly delayed compared to the observations especially for the TimesFM+Cov prediction. Furthermore, the maximum enhancement after the storm onset seems to be underestimated by the TimesFM+Cov, as also seen in Figure~\ref{fig:ltime-pred-error}e. This underestimation is also seen at L = 4.5 and 5.5. 

\subsection{Long forecast horizon with TimesFM+Cov} \label{subsec:results-timesfm} 

As shown in Section~\ref{subsec:results-6h}, we found that the TimesFM+Cov provides the best overall results quantitatively and qualitatively. In this Section, we experiment to see whether the TimesFM+Cov could provide quality long-horizon forecasts. Without any new fine-tuning on the ridge regression part (equation~(\ref{eq:rss-ridge})), we perform the inference (i.e., prediction) with the past dynamic covariates in equation~(\ref{eq:ridge-reg}) and the TimesFM on the residuals in equation~(\ref{eq:res-proj}) in order to obtain the final prediction in equation~(\ref{eq:tfm-cov-final}). The only change here is the time horizon $H$ that is extended up to 72 hours in advance. Furthermore, we also perform the prediction for lower time resolutions of 12h and 24h of the data. 

Table~\ref{table:long-horizon-all} shows the average $R^2$ score over all L-shells from the TimesFM+Cov for the forecast horizons of 6, 12, 24, 48, and 72 hours. For all temporal resolutions of the data, we find that the $R^2$ score decreases for the longer time horizons. The decrease is most pronounced for the 6h data resolution where the $R^2$ score decreases from 0.91 for 6h forecast to -0.48 for 72h forecast. For the 12h data resolution, the $R^2$ score decreases from 0.85 for 12h forecast to 0.19 for 72h forecast. Lastly, for the 24h data resolution, the $R^2$ score decreases from 0.82 for 24h forecast to 0.48 for 72h forecast. This shows that the multi-step forecasting performs rather poorly for the TimesFM+Cov. Furthermore, we find that for the 72h forecast, the 3-step forecasting of the 24h data performs much better than the 6-step forecasting of the 6h forecast. Therefore, to provide long-horizon forecasts up to three days, one should limit to the data resolution of 24h. 

\begin{table}[h]
\caption{Average $R^2$ score of the TimesFM+Cov over all L-shells for the different forecast horizons and different temporal resolutions.} \label{table:long-horizon-all}
\centering 
\begin{tabular}{p{1.8cm}p{1.8cm}p{1.8cm}p{1.8cm}p{1.8cm}p{1.8cm}}
\hline 
Temporal resolution & 6h horizon & 12h horizon & 24h horizon & 48h horizon & 72h horizon \\
\hline 
6h             & 0.91         & 0.79          & 0.58          & 0.26          & -0.48          \\
12h            & -            & 0.85          & 0.69          & 0.41          & 0.19          \\
24h            & -            & -             & 0.82          & 0.52          & 0.48      \\   
\hline 
\end{tabular}
\end{table}


\section{Discussion}\label{sec:discussions}

We developed a pipeline for 1-MeV electron flux forecasts in the outer radiation belt as observed by SEM-2 onboard POES NOAA-15. The pipeline is inspired by PreMevE-2.0 but with different datasets for the target 1-MeV electron flux, the input near 1-MeV electron flux at GEO, and slightly different sets of L1 observations and geomagnetic indices. The purpose is to develop a pipeline that can be continuously updated with new observations, thanks to the usage of SEM-2 electron flux observations onboard of ongoing POES, MetOp, and GOES, among others. In terms of ML algorithms, we include a Transformer-based model ``TransEncode" in addition to the LSTM, CNN, and linear regression used in PreMevE-2.0. Particularly, we also employ a pre-trained timeseries model ``TimesFM" using its default and its hybrid applications. We find that the hybrid application of TimesFM, the TimesFM+Cov, yields the best results surpassing all the algorithms for the test data in January - June 2024. In the following, we discuss how our best model performs compared to the others and to some existing works. 

Our main results reported in Section~\ref{sec:results} mainly concern the 6h forecast (Section~\ref{subsec:results-6h}). We first compared the $R^2$ scores over all L-shells for all the developed models (Figure~\ref{fig:r2_lshells_6h}) before showing the qualitative results among the top performing models (Figure~\ref{fig:ltime-pred-error}). The best average $R^2$ score of 0.90 over all L-shells was found for the TimesFM+Cov model, while the average $R^2$ scores of 0.72 - 0.78 were found for other models (linear regression, TransEncode, Conv1D, and LSTM). Due to the lack of studies for the same forecasting horizon and for the same period of out-of-sample test data, we cannot directly compare our results to the existing literature. Nevertheless, we may mention some results from PreMevE \cite{2019SpWea..17..438C} here as they show results for the 5h forecast. The prediction efficiency (i.e., $R^2$) for their best SubModel2 was found to be 0.9 at L=3.5 and 0.5 at L=6.0 for the test data in December 2015 to August 2016. We note that their test data period is during the solar declining phase of the solar cycle 24 while ours is during the solar ascending phase of the solar cycle 25 that includes the strongest geomagnetic storm since 2003 (10-12 May 2024 event). To put the performance of our best model to the context, we found the $R^2$ score of 0.95 at L=3.5 and 0.76 at L=6.0. In other words, for our best model, the performance is better especially at the high L. For our second best models for these L values, we found $R^2$ score of 0.9 at L=3.5 (Conv1D) and $R^2$ of 0.52 at L=6.0 (LSTM), relatively similar to the values found in PreMevE.

The results described above imply that our ML models excluding TimesFM+Cov provide rather similar model performance for L=3.5 and L=6.0. In PreMevE-2.0, they showed that the linear regression model is often the most successful when compared to other models especially for L $\leq$ 4.8. This suggests that the relationship between dynamics of trapped 1-MeV electrons and the precipitating electrons is dominated by linear components \cite{2020SpWea..1802399P}. In Figure~\ref{fig:r2_lshells_6h}, we generally find poorer results for higher L-shells for all models except the TimesFM+Cov. This is likely because the different physical mechanisms operate at the inner L-shells and the upper L-shells. It was suggested that for larger L-shells, it is dominated by physical processes other than wave-particle interaction such as radial diffusion and adiabatic effects \cite{2019SpWea..17..438C}. Additionally, \cite{2015SpWea..13..853S} found that the energetic electron flux for the innermost region of the outer radiation belt is best predicted by using the Dst index, while it is best described by the solar wind velocity and the dynamic pressure for the central (L $\geq$ 4.8) and outermost (L $\geq$ 5.6) regions, respectively. In our case, the $R^2$ scores drop to under 0.8 from L=4.8 onward for all models except the TimesFM+Cov that shows the $R^2$ scores under 0.9 from this L. Furthermore, when using the TimesFM+Cov, we find that the difference $R^2$ scores between the inner and outer belt regions are less pronounced. 

We may discuss on why the TimesFM+Cov works better than other models. When using the TimesFM zero-shot for an autoregression of the 1-MeV electron flux variation, it shows the worst results with the average $R^2$ of 0.59, even worse than the persistence baseline. This shows that the default application of TimesFM cannot be used with our problem. When using the TimesFM+Cov, we find the best result with the average $R^2$ of 0.90. Indeed, the ridge regression step first fits the linear variation between the past dynamic covariates of all the input variables. Then, the TimesFM is applied to forecast the residuals which now contains principally the nonlinear components. These suggest that the TimesFM+Cov predicts both linear and nonlinear variations of the 1-MeV electron flux, which may explains why it outperforms other models. For other models, in contrast, we train them from scratch while fitting them to both linear and nonlinear components simultaneously. This may explains why the classic models do not work as well as the TimesFM+Cov that handled the linear and nonlinear parts separately. 

Overall, the TimesFM+Cov shows significant improvement in terms of the model performance for all L-shells, and especially at the high L-shells. Comparing to our second-best models, the improvements in $R^2$ scores are found to be about 12\% at L=2.8 (compared to Conv1D), about 5\% at L=3.6 (compared to Conv1D), and about 50\% at L=6.0 (compared to LSTM). Qualitatively, the TimesFM+Cov model also shows best results with the correctly reproduced storm onsets and storm variations (Figure~\ref{fig:ltime-pred-error}). However, the TimesFM+Cov model somewhat underestimates the storm peak intensities (with about 10-20\% on the percentage error) and yields some delays (up to 24 hours at L=3.5, but less for higher L shells) on the storm onsets. We aim to investigate whey such delays are present and how can we minimize the delays for better storm onset predictions in the future work. Additionally, we investigate how the TimesFM+Cov model performs for longer forecast horizons up to 3 days in advance. As shown in Table~\ref{table:long-horizon-all}, it turns out that the TimesFM+Cov performs less well for longer forecast horizons especially for the 6h temporal resolution. We note that the current TimesFM+Cov is mostly optimized for the 6h data resolution and with 1-step forecasting. To obtain better performance for the longer forecast horizons, our results suggest that we need to fine-tune the TimesFM+Cov (i.e., finding an optimal context length, fitting the ridge regression) for each time resolution and perhaps each forecast horizon. This step is left for future work. 

In conclusions, we find that the TimesFM+Cov model is most promising for forecasting the 1-MeV electron flux in the outer belts as observed by POES P6. We note that, unlike our classic ML models which are trained from scratch using 11-years of data, the TimesFM+Cov model takes only 6 months of the most recent data to fine-tune the ridge regression step. This aspect is rather important considering that most satellite missions in the radiation belt environments do not have long datasets like those of POES. We note, nevertheless, that our usage of TimesFM+Cov should be further optimized in the future to make the model more operational. For instance, as mentioned in Section~\ref{subsubsec:tfm-cov}, we use more input variables for the TimesFM+Cov as motivated by the definition of the covariates rather than using reduced sets of variables as done for other ML models, which are motivated by the inputs similar to PreMevE-2.0. Indeed, we should prefer a lower number of input variables as certain observations may not be available. These aspects should be studied in the future in order to make the TimesFM+Cov more operational.


\section{Conclusions and Perspectives}\label{sec:conclusions}

We developed a prototype ML-based pipeline for 1-MeV electron flux forecasts in the outer radiation belt (2.8 $\leq$ L $\leq$ 6.0) with a horizon forecast from 6 hours to 3 days. Inspired by PreMeV-E 2.0, we first developed classic ML techniques including linear regression, neural network algorithms (Conv1D, LSTM), and a Transformer-Encoder. Unlike PreMeVE-model series, we focus on the long-term observations of 1-MeV electron flux by SEM-2 onboard POES NOAA-15 as the forecast target. We also applied for the first time a pre-trained timeseries foundation model ``TimesFM" using its default zero-shot application and its hybrid application with the past dynamic covariates. The purpose is to develop a high-fidelity modeling framework toward a more-operational approach for energetic electron flux variation forecasting in the outer Earth's radiation belt.  

Using data in 2013 - 2023, we first trained classic ML techniques and evaluate their results with the out-of-sample test data in January - June 2024 including the extreme geomagnetic storm on 10-12 May 2024. Using 6-month data in late 2023 to fine-tune the hybrid application of TimesFM (TimesFM+Cov), we compare their results to the classic ML models on the same test data. We find that the TimesFM+Cov provides the best results with an $R^2$ score above 0.9 for L $\leq$ 4.7 and it remains above 0.75 for L = 6.0, outperforming all the classic models especially at the high L shells. Compared to the second best models, the TimesFM+Cov yields an improvement of $R^2$ score for about 12\% at the lowest L shell and about 50\% at the highest L shell compared to the Conv1D and LSTM models, respectively. Overall, the TimesFM+Cov shows good qualitative results similar to the observations although with some delays on the geomagnetic storm onsets (up to about a day at L=3.5) and some underestimated peaks during the geomagnetic storms (5-10\% in log flux). Future work should be focused on evaluating the TimesFM+Cov results in more detail and optimize it for better accuracy for longer forecast horizons.

To our knowledge, it is the first time that a pre-trained timeseries foundation model is adapted for energetic electron forecasting in the Earth's outer radiation belt. By performing a ridge (linear) regression with the past dynamic covariates first and then applying the TimesFM on the residuals, we demonstrate that this hybrid approach can outperform the classic supervised ML training approach from scratch while using less data (6 months instead of 11 years in our case). Yet, we should further develop and exploit this approach to better understand better it could model more precisely both linear and nonlinear processes of energetic electron acceleration in the outer belts. Our current work offers a proof-of-concept that would serve as a base framework toward more operational approach of outer belt energetic electron forecasting with longer time horizons.

\section*{Open Research Section}
All of the used data are publicly available and can be accessed from the following sources. The L1b POES data can be accessed from \url{https://www.ncei.noaa.gov/data/poes-metop-space-environment-monitor/access/l1b/v01r00/}. The L2 data from GOES 15 are accessible from \url{https://www.ncei.noaa.gov/data/goes-space-environment-monitor/access/avg/} and those of GOES 17 and 18 data are accessible from \url{https://data.ngdc.noaa.gov/platforms/solar-space-observing-satellites/goes/}. All the solar wind and IMF parameters, as well as the geomagnetic indices can be accessed from the Coordinated Data Analysis Web (CDAWeb; \url{https://cdaweb.gsfc.nasa.gov/}). This research used version 6.1.1 \cite{sunpy_community2020} of the SunPy open source software package (\url{https://github.com/sunpy/sunpy/tree/v6.1.1}); version 2.3.1 of the PyTorch deep learning library \cite{2019arXiv191201703P}, and the version 2.0 of the TimesFM model (\url{https://huggingface.co/google/timesfm-2.0-500m-jax}). 

\section*{Acknowledgments}

This research work is conducted at Research Institute in Astrophysics and Planetology (IRAP), University of Toulouse, in the framework of the Augura Space project supported by the French Institute for Research in Computer Science and Automation (INRIA). We acknowledge the funding from INRIA, and the French national space agency (CNES). The authors are grateful to the CLEPS infrastructure from the INRIA of Paris for providing resources and support. We appreciate the discussion with Robert Ecoffet and his team at CNES. We thanks for the feedback from Jean-François Ripoll at the French Alternative Energies and Atomic Energy Commission (CEA) Paris-Saclay, France. 

\bibliographystyle{unsrt}  
\bibliography{references}  

%
%
%
%

\end{document}